# Superconductivity enhancement in the S-doped Weyl semimetal candidate MoTe$_2$

F. C. Chen[1,2], X. Luo[1*], R. C. Xiao[1,2], W.J. Lu[1], B. Zhang[5], H.X. Yang[5], J. Q. Li[5,6],

Q. L. Pei[1], D. F. Shao[1], R. R. Zhang[3], L. S. Ling[3], C. Y. Xi[3], W. H. Song[1] and Y. P. Sun[3,1,4*]

[1] Key Laboratory of Materials Physics, Institute of Solid State Physics, Chinese Academy of Sciences, Hefei, 230031, China

[2] University of Science and Technology of China, Hefei, 230026, China

[3] High Magnetic Field Laboratory, Chinese Academy of Sciences, Hefei, 230031, China

[4] Collaborative Innovation Center of Advanced Microstructures, Nanjing University, Nanjing, 210093, China

[5] Beijing National Laboratory for Condensed Matter Physics, Institute of Physics, Chinese Academy of Sciences, Beijing, 100190, China

[6] Collaborative Innovation Center of Quantum Matter, Beijing, 100190, China

## Abstract

Two-dimensional (2D) transition-metal dichalcogenide (TMDs) MoTe$_2$ has attracted much attention due to its predicted Weyl semimetal (WSM) state and a quantum spin Hall insulator in bulk and monolayer form, respectively. We find that the superconductivity in MoTe$_2$ single crystal can be much enhanced by the partial substitution of the Te ions by the S ones. The maximum of the superconducting temperature $T_C$ of MoTe$_{1.8}$S$_{0.2}$ single crystal is about 1.3 K. Compared with the parent MoTe$_2$ single crystal ($T_C$=0.1 K), nearly 13-fold in $T_C$ is improved in MoTe$_{1.8}$S$_{0.2}$ one. The superconductivity has been investigated by the resistivity and magnetization measurements. MoTe$_{2-x}$S$_x$ single crystals belong to weak coupling superconductors and the improvement of the superconductivity may be related to the enhanced electron-phonon coupling induced by the S-ion substitution. A dome-shape superconducting phase diagram is obtained in the S-doped MoTe$_2$ single crystals. MoTe$_{2-x}$S$_x$ materials may provide a new platform for our understanding of superconductivity phenomena and topological physics in TMDs.

* Corresponding author: xluo@issp.ac.cn and ypsun@issp.ac.cn

# I Introduction

Layered transition-metal dichalcogenides (TMDs) materials, named as $MX_2$, where M is a transition metal (Ta, Mo and W) and X is a chalcogen (S, Se and Te), have attracted renewed interest owing to their rich physical properties and promising potential applications.[1-10] For example, artificial van der Waals heterostructures with high on/off current ratio in TMDs are promising as an active channel for next-generation devices.[2] Topological field-effect transitions based on quantum spin Hall (QSH) insulators are novel type of device.[9] At the same time, charge-density waves and superconductivity have also been observed in TMDs.[10] Particularly, a dome-shaped superconducting phase diagram is observed in a gate-tuned $MoS_2$ device and the $WTe_2$ bulk crystal driven by the pressure.[4-6]

Recently, $MoTe_2$ has been attracted much attention because it was predicted a new type-II Weyl semimetal (WSM) candidate.[11] $MoTe_2$ crystalizes into three different phases: 2H, 1T' and $T_d$ ones. The 2H phase shows a semiconducting behavior. On the other hand, the 1T' and $T_d$ phases present semi-metallic and exhibit pseudo-hexagonal layers with zig-zag metal chains. The $T_d$ compound can be obtained by cooling the 1T' phase down to 245 K.[8, 12] The $T_d$ phase of $MoTe_2$, which is isostructural to $WTe_2$, is predicted to be a candidate WSM.[13] Very recently, the $T_d$-$MoTe_2$ is reported to be a superconductivity with $T_C$=0.1 K, the small applied pressure can dramatically enhance the $T_C$ and a dome-shaped superconducting phase diagram under the applied pressure is observed.[8] The $MoTe_2$ opens a door to study the interaction of topological physics and superconductivity in the bulk materials. Provoked by above reported work, since the superconductivity can be much enhanced by the applied pressure in $MoTe_2$ bulk crystal, which implies that the $T_d$-$MoTe_2$ may be also sensitive to the chemical pressure induced by the ion substitution. In this work, we did the partial substitution of Te ions by S ones in $MoTe_2$ single crystal. The enhanced superconducting temperature $T_C$ is observed and the maximum of the $T_C$ in $MoTe_{1.8}S_{0.2}$ single crystals about 1.3 K. A dome-shape superconducting phase diagram is obtained in $MoTe_{2-x}S_x$ single crystals and the possible reasons of the enhanced superconductivity have been discussed. The $MoTe_{2-x}S_x$ materials may provide a new platform to understand the superconductivity phenomena and topological physics in TMDs.

# II Experimental details

MoTe$_2$ and MoTe$_{2-x}$S$_x$(0⩽x⩽1) single crystals were grown by the chemical vapor transport method using polycrystalline MoTe$_2$ and MoTe$_{2-x}$S$_x$ as raw materials and I$_2$ as a transport agent. Firstly, we made the polycrystalline MoTe$_2$ and MoTe$_{2-x}$S$_x$(0⩽x⩽1). Mo (Alfa Aesar, 99.9 %) and Te or S(Alfa Aesar, 99.9 %) powders with a stoichiometric ratio were ground, pressed into pellets, and put in an evacuated quartz tube. All were done in an Ar-filled glove box. The sealed quartz tube was heated to 800 ºC for 20 hours and kept for 7 days, then quenched to ice water. The polycrystalline MoTe$_2$ or MoTe$_{2-x}$S$_x$ and the agent I$_2$ were mixed and sealed into another evacuated quartz tubes. The sealed quartz tubes were put in a two-zone tube furnace. The crystal growth recipe is followed the reported paper.[12] The hot side is about 1000 ºC and the cold side is 900 ºC, and dwelled for 7 days. In order to improve the quality of single crystals, we quenched the quartz tubes into ice-water as soon as possible before the growth sequence ending. Plate-like shape single crystals with shinning surfaces were obtained. The size of the crystal was about 4*4*0.5 mm$^3$. The single crystals were air-stable and can be easily exfoliated. Powder X-ray diffraction (XRD) patterns were taken with Cu $K_{\alpha1}$ radiation (λ=0.15406 nm) using a PANalytical X'pert diffractometer at room temperature. The element analysis of the single crystals was performed using a commercial energy dispersive spectroscopy (EDS) microprobe. The element compositions of the single crystals used in the text are the ones obtained from EDS measurements. The magnetic properties were carried out by the magnetic property measurement system with a $^3$He cryostat (MPMS-XL7). The electrical transport measurements were performed in a $^4$He cryostat from 300 K to 2 K, and in a $^3$He cryostat down to 0.35 K by a four-probe method to eliminate the contact resistance. The measurement of specific heat was carried out by a heat-pulse relaxation method on Physical Properties Measurement System (PPMS-9T). Raman spectra were measured on Horiba T64000, with excitation wavelength 647.4 nm and the power density was kept below 20 mW cm$^{-2}$ in order to minimize the heating effects. The density functional theory (DFT) calculations were carried out using QUANTUM ESPRESSO package [14] with ultrasoft pseudopotentials. The exchange-correlation interaction was treated with the generalized gradient approximation (GGA) with Perdew-Burke-Ernzerh (PBE) of parametrization.[15] The energy cutoff for the plane-wave basis set was 65 Ry. The Vanderbilt-Marzari Fermi smearing method with a smearing parameter of $\sigma = 0.02$ Ry was used for the calculations of the total energy and electron charge density. The van der Waals interactions were treated by a semiempirical dispersion correction scheme (DFT-D).[16, 17] The crystal structures were optimized with respect to lattice parameters and atomic positions. Brillouin zone sampling is performed on the Monkhorst-Pack (MP) meshs [18] of $6 \times 12 \times 4$ for structure relaxation and $12 \times 24 \times 8$ for density of states (DOS) calculation. Electron diffraction and scanning transmission electron microscopy (STEM) experiments were performed in the JEOL ARM200F equipped with double aberration correctors and cold field emission gun operated at 200kV.

## III Results and Discussion

MoTe$_2$ crystallizes into three different phases, as shown in Fig. 1 (a)-(c). In 2H phase, the Mo atom has trigonal prismatic coordination with the Te atoms. The 1T' phase (space group *P2$_1$/m*), which is also called *β*-phase, is a monoclinic lattice and is stable at room temperature. The T$_d$ phase (space group *Pmn2$_1$*) is an orthorhombic lattice and the 1T' one is slight sliding of layer-stacking of the T$_d$ phase. We selected MoTe$_{1.8}$S$_{0.2}$ single crystal as a typical sample to do the detailed experiments. Figure 1 (d) shows the X-ray patterns of MoTe$_{1.8}$S$_{0.2}$ single crystals along [001] direction. We did the powder X-ray of crushed MoTe$_{1.8}$S$_{0.2}$ single crystals. However, we found that the X-ray data can be well fitted based on the orthorhombic or monoclinic symmetry, as shown in Fig. S1. It is difficult for us to distinguish the structure of MoTe$_{1.8}$S$_{0.2}$ single crystal at the room temperature. We also carried out TEM characterization of the MoTe$_{1.8}$S$_{0.2}$ single crystal. As shown in Fig. 1 (e), the diffraction spots on the [001] selected area electron diffraction (SAED) pattern can be well indexed by the orthorhombic or monoclinic symmetry. However, the SAED pattern (Fig. 1(f)) and annular bright field (ABF) STEM image (Fig. 1(g)) taken along the [010] zone axis direction demonstrate evident structure difference comparing with the previous reported orthorhombic or monoclinic structure, it is recognizable that the MoTe(S)X$_6$ octahedral are much less distorted.

In order to compare the lattice parameters of MoTe$_2$ and MoTe$_{1.8}$S$_{0.2}$ single crystals, we also did the XRD experiment of the crushed MoTe$_2$ single crystals. Figure S2 shows the experimental and calculated XRD patterns. The calculated lattice parameters of crushed MoTe$_2$ and MoTe$_{1.8}$S$_{0.2}$ single crystals and the reported MoTe$_2$ one are summarized in Table I of the supporting materials. In addition, in order to get more some structural information on the MoTe$_{2-x}$S$_x$ single crystal, we also did the Raman experiments. As shown in Fig. S3, except for MoTeS single crystal (changing into 2H phase), all another crystals show similar Raman signals at room temperature. The temperature dependence of Raman spectra of the MoTe$_2$ single crystal is shown in Fig. S4, it seems that it is even difficult to tell the phase transition from present Raman experiments. That means we still can't tell the structure of S-doped MoTe$_2$ crystals. Thus, more detailed experiments are really needed to perform in the future.

Figure 2 shows the evolution of the resistance as a function of temperature $\rho(T)$ in a MoTe$_{1.8}$S$_{0.2}$ single crystal down to 0.35 K. For the MoTe$_2$ single crystal, we did the measurement down to 2 K under the cooling and warming modes. As shown in Fig. 2, an anomaly with a hysteresis in the $\rho(T)$ of MoTe$_2$ single crystals observed, which is associated with the first-order structural phase transition from 1T' phase to T$_d$ phase around 240 K and consistent with the reported results.[8,12] We applied the Fermi-liquid model $\rho(T)=\rho_0+A*T^2$, which $\rho_0$ and $A$ are the residual resistivity and a constant, respectively, to the curve below 50 K as shown in the left inset of Fig. 2. The model analysis yielded the residual resistivity $\rho_0$ of $1.36*10^{-4}$ $\Omega$ cm and the coefficient A of 0.00238 μΩ cm, indicating Fermi-liquid-like behavior for MoTe$_2$ single crystal. For MoTe$_{1.8}$S$_{0.2}$ single crystal, as shown in the right inset of Fig. 2, a clear zero-resistivity behavior is observed around 1.3 K. We measured the $\rho(T)$ under applied magnetic field H=0.5 T, the zero-resistivity temperature moves to 0.7 K, which indicates the superconductivity may occurs in MoTe$_{1.8}$S$_{0.2}$ single crystal. On the other hand, we also find that there exists the minimum of the resistivity around $T_{min}$=8 K before the superconducting transition in MoTe$_{1.8}$S$_{0.2}$ single crystal, which is similar to the transport properties of the WTe$_2$ single crystal under the applied high pressure.[5] The presence of the minimum of resistivity may be related to the structural abnormalities induced by the substitution of Te ions by S ones in MoTe$_{1.8}$S$_{0.2}$ single crystal.

To test the superconductivity of MoTe$_{1.8}$S$_{0.2}$ single crystal further, we did the magnetic measurements. Figure 3 presents the temperature dependence of magnetization $M(T)$ with the zero-field and field cooling (ZFC and FC)modes under an applied magnetic field of 10 Oe. The diamagnetic signal, in other words, the superconducting transition, can be obviously seen at $T_C$=1.3 K in the left inset of Fig. 3, which is in good agreement with the resistance measurements. The calculated shielding superconducting volume fraction at 0.5 K is about 60 %, which means the bulk superconducting behavior of the MoTe$_{1.8}$S$_{0.2}$ single crystal. The magnetization as a function of magnetic fields at 0.5 K is shown in the right inset of Fig. 3. The clear hysteresis is observed, which indicates a typical type-II superconducting behavior of MoTe$_{1.8}$S$_{0.2}$ single crystals. Compared with the parent MoTe$_2$ ($T_C$=0.1 K), nearly 13-fold of the superconducting temperature is observed in S doped MoTe$_{1.8}$S$_{0.2}$ single crystal.

The temperature dependence of specific heat $C_P$ also provides more information about the normal state properties. Figure 4 shows the temperature dependence of the heat capacity $C_P$ of MoTe$_2$ and MoTe$_{1.8}$S$_{0.2}$ single crystals at $H$= 0 above $T_C$. $C_P/T$ varies almost linearly with $T^2$. The Sommerfeld constant $\gamma_N$ is obtained from the fit $C_P/T = \gamma_N + \beta T^2$ ($\gamma_N$ and $\beta$ are $T$-independent coefficients), where $\gamma_N$ is the normal-state electronic contribution and $\beta$ is the lattice contribution to the specific heat. The fitting results yield $\gamma_N$ = 3.06mJ mol$^{-1}$ K$^{-2}$ and $\beta$= 0.758mJ mol$^{-1}$ K$^{-4}$ for MoTe$_2$ single crystal and $\gamma_N$ = 2.07mJ mol$^{-1}$ K$^{-2}$ and $\beta$= 0.635mJ mol$^{-1}$ K$^{-4}$ for MoTe$_{1.8}$S$_{0.2}$ one, respectively. The resulted DOS at Fermi surface $N(E_F)$ are 1.33 states/eV and 0.88 states/eV for MoTe$_2$ and MoTe$_{1.8}$S$_{0.2}$ single crystals, respectively. The Debye temperature $\Theta_D$ can be determined from the coefficient of the $T^2$ term $\beta=N(12/5)\pi^4 R\Theta_D^{-3}$, where $R$= 8.314 J mol$^{-1}$ K$^{-1}$ and $N$= 3 for the MoTe$_2$ and MoTe$_{1.8}$S$_{0.2}$ single crystals. The Debye temperatures $\Theta_D$ are 135 K and 143 K for MoTe$_2$ and MoTe$_{1.8}$S$_{0.2}$ single crystals, respectively. All the fitting results are summarized in **Table I**. An estimation of the strength of the electron-phonon coupling (EPC) can be derived from the McMillan formula: [19, 20]

$$\lambda_{ep} = \frac{\mu^* \ln\left(\frac{1.45 T_C}{\Theta_D}\right) - 1.04}{1.04 + \ln\left(\frac{1.45 T_C}{\Theta_D}\right)(1 - 0.62\mu^*)} \quad . \tag{1}$$

By assuming the Coulomb pseudopotential $\mu^*$=0.1, the EPC constant $\lambda_{ep}$ is estimated to be 0.32 and 0.49 for MoTe$_2$ and MoTe$_{1.8}$S$_{0.2}$ single crystals, respectively. The calculated EPC show that both of MoTe$_2$ and MoTe$_{1.8}$S$_{0.2}$ single crystals are weak coupling superconductors.[19] Because the $N(E_F)$ of MoTe$_2$ single crystal is larger than that of MoTe$_{1.8}$S$_{0.2}$ one, the improvement of the superconductivity in MoTe$_{1.8}$S$_{0.2}$ single crystal can't be explained according to the variation of DOS of Fermi surface $N(E_F)$ and it may be related to the enhancement of the EPC by the S-ion substitution.

In order to study the superconducting evolution with S-doping level, we also grew and studied the electrical transport properties of the MoTe$_{2-x}$S$_x$(0≤x≤1) single crystals. All the zero-resistivity temperature $T_C$, the minimum temperature of the resistivity $T_{min}$ and the structural phase transition temperature $T_S$ vs. the doped S contents $x$ are summarized in the $T$-$x$ phase diagram shown in Fig. 5. The structural phase transition temperature $T_S$ is defined from the $\rho(T)$ data and similar to that of MoTe$_2$ single crystal,[12] as shown in the inset of Fig. S5 (a). Note that the $T_S$ anomaly slowly increases with the little S-doped content $x$, then decreases rapidly with the increasing $x$. Further, the $T_S$ disappears at $x=0.18$. For the superconducting temperature $T_C$, it increases is nearly linearly up to a maximum $T_C=1.3$ at $x=0.2$, as shown in Fig. S6. From this maximum, the $T_C$ begins to decrease with increasing $x$, with the transition eventually disappearing at $x=0.37$. Thus, a dome-shaped superconducting phase diagram is obtained for MoTe$_{2-x}$S$_x$ single crystal. For the S doped MoTe$_{2-x}$S$_x$crystals, the lower S-doping at Te sites can sharply increase in $T_C$, which is concomitant with enhancement of structural transition. Then, the $T_C$ increases still slowly, however, the $T_S$ is suppressed and disappearing with the increasing $x$, which are similar to with the results from the applied pressure in MoTe$_2$ single crystal.[8] On the other hand, when the content $x$ is larger than 0.18, which is in accord with the content where the structural phase disappears, the minimum of resistivity $T_{min}$ presents and increases slowly with the $x$, then tends to a constant value when the $x$ is larger or equal to 0.37. For $x$ is equal to 1, the crystal presents 2H phase and shows a semiconducting behavior, as shown in Fig. S3 and S5. However, we did not get the single crystals when $x$ is larger than 1, which means the solution limit of S ions may be $x=1$ in MoTe$_{2-x}$S$_x$ single crystals.

To understand the enhanced superconductivity in MoTe$_{2-x}$S$_x$ single crystals, we did the first principle calculations. In order to simulate the S-doping effect, we substitute one Te atom in MoTe$_2$ unit cell by one S atom, i.e., the doped MoTe$_{7/4}$S$_{1/4}$ was theoretically considered. The results show that the GGA slightly overe stimatesthe lattice parameters of MoTe$_2$ with 1T' and T$_d$ structures compared with those obtained from experiments (see **Table I** in the supporting materials, the difference is less than 2%). For the MoTe$_{7/4}$S$_{1/4}$ sample, we test the cases that S atom locates at four inequivalent Te-sites in the unit cell respectively. By comparing the total energies, we found both in 1T' and T$_d$ structure, S atom tends to replace the Te atom for forming the shortest Mo-Te bond. As expected, the S subtitution makes the structures shrinking. Figure S7 shows the calculated DOS for MoTe$_2$ and MoTe$_{7/4}$S$_{1/4}$ with 1T' and T$_d$ structures. For the MoTe$_2$, our calculation is well consistent with previous calculations.[21] One can note that in both structures S doping rarely changes the overall structure of DOS. S doping slightly increases the band widths, which should be due to the enhaced $p-d$ hybridization in the shrinked structures. In both structures the DOS at $E_F$ ($N(E_F)$) decreases upon S doping, as shown in Fig. S5 (c), which is good agreement with the results from the $C_P$ measurements. That is, the value of Sommerfeld constant $\gamma_N$ of MoTe$_{1.8}$S$_{0.2}$ crystal is lower than that of MoTe$_2$ one. Therefore, the decreased $N(E_F)$ does not support the S-doping induced superconductivity ehancement in 1T' and T$_d$ structures and consistent with the results from the experimental results. The real structure of our experimentally prepared MoTe$_{2-x}$S$_x$ crystals might be different from 1T' and T$_d$, which needs to be further investigated both in experment and theory.

Now, we pay attention to the EPC strength $\lambda_{ep}$. As we known, the EPC strength $\lambda_{ep}$ for a material can be qualitatively expressed as:

$$\lambda_{ep} = \sum_\alpha \frac{\langle I_\alpha^2 \rangle N_\alpha(E_F)}{M_\alpha \langle \omega_\alpha^2 \rangle}, \tag{2}$$

where the $\langle I_\alpha^2 \rangle, M_\alpha$, and $\langle \omega_\alpha^2 \rangle$ are the mean square EPC matrix element averaged over Fermi surface, atomic mass, and averaged squared phonon frequency of the $\alpha$th atom in the unit cell, respectively.[22-24] We can simply compare the variations of $\langle I_\alpha^2 \rangle, M_\alpha, \langle \omega_\alpha^2 \rangle$ and $N(E_F)$ of MoTe$_2$ and MoTe$_{1.8}$S$_{0.2}$ crystals. As we know, the S ions are lighter than Te one, so the $M_\alpha$ of MoTe$_{1.8}$S$_{0.2}$ single crystal (332.03 g/mol) is smaller than that of MoTe$_2$ one (351.14 g/mol). Meanwhile, the Debye temperature $\Theta_D$ is in proportion to the phonon frequency $\langle \omega_\alpha^2 \rangle$ and that of MoTe$_{1.8}$S$_{0.2}$ crystal ($\Theta_D$=143 K) is little larger than that of MoTe$_2$ one ($\Theta_D$=135 K). So the denominators $M_\alpha \langle \omega_\alpha^2 \rangle$ of MoTe$_2$ and MoTe$_{1.8}$S$_{0.2}$ crystals may be close. Let's pay attention to the numerator, the $N(F)$ of MoTe$_{1.8}$S$_{0.2}$ single crystal ($N(F)$=0.88 states/eV) is just 67 % of that of MoTe$_2$ one ($N(F)$=1.33 states/eV), which indicates the mean square EPC matrix element averaged over Fermi surface $\langle I_\alpha^2 \rangle$ of MoTe$_{1.8}$S$_{0.2}$ single crystal should be much larger than that of MoTe$_2$ one because the ECP strength $\lambda_{ep}$ of MoTe$_{1.8}$S$_{0.2}$ ($\lambda_{ep}$=0.49) is over one and half times than that of MoTe$_2$ ($\lambda_{ep}$=0.32). Thus, we can propose that the improvement of superconductivity in MoTe$_{1.8}$S$_{0.2}$ single crystal may be related to the dramatic changing or reconstructing of the Fermi surface in the new structure induced by the substitution of Te ions by S one. However, the compressive exploration of superconductivity in MoTe$_{2-x}$S$_x$ from both experimental and theoretical perspectives is needed in the future.

## IV Conclusion

We find that the superconductivity in MoTe$_2$ single crystal can be much enhanced by the partial substitution of the Te ions by the S ones. The maximum of the superconducting temperature $T_C$ of MoTe$_{1.8}$S$_{0.2}$ single crystal is about 1.3 K. Compared with the parent MoTe$_2$ single crystal ($T_C$=0.1 K), nearly 13-fold in $T_C$ is improved in Mo Te$_{1.8}$S$_{0.2}$ one. The superconductivity has been investigated by the resistivity and magnetization measurements. MoTe$_{2-x}$S$_x$ single crystals belong to weak coupling superconductors and the improvement of the superconductivity may be due to the enhanced EPC induced by the S-ion substitution. A dome-shape superconducting phase diagram is obtained in the S-doped MoTe$_2$ single crystals. The MoTe$_{2-x}$S$_x$ materials may provide a new platform to understand the interplay between the superconductivity and topological physics in TMDs.

## Acknowledgement

This work was supported by the Joint Funds of the National Natural Science Foundation of China and the Chinese Academy of Sciences' Large-Scale Scientific Facility under contracts (U1432139, U1232139), the National Nature Science Foundation of China under contracts (51171177, 11404342), the National Key Basic Research under contract 2011CBA00111, and the Nature Science Foundation of Anhui Province under contract 1508085ME103, 1408085MA11.

**Author contributions:**

Y.S. proposed the work. X.L. and Y.S. designed the research. F.C. grew the single crystals. F.C. and Q.P. made the material characterizations with the help of L.L., C.X. and W.S.; R.X., D.S. and W.L. did the theoretic calculations. R.Z. carried out the Raman measurements. B. Z., H. Y. and J. L. performed the TEM measurements. F.C., X.L. and Y.S. co-wrote the paper. All authors commented to the manuscript.

# References:

[1] Dong HoonKeum, Suyeon Cho, Jung Ho Kim, Duk-Hyun Choe, Ha-Jun Sung, Min Kan, Haeyong Kang, Jea-Yeol Hwang, Sung Wng Kim, Heejun Yang, K. J. Chang, and Young Hee Lee, *Nature Physics* **11**, 482 (2015).

[2] Suyeon Cho, Sera Kim, Jung Ho Kim, Jiong Zhao, JinbongSeok, Dong HoonKeum, JaeyoonBaik, Duk-Hyun Choe, K. J. Chang, KazuSeunaga, Sung Wng Kim, Yong Hee Lee, Heejun Yang, *Science*, **349**, 625 (2015).

[3] Mazhar N. Ali, Jun Xiong, Steven Flynn, Jing Tao, Quinn D. Gibson, Leslie M. Schoop, Tian Liang, Neel Haldolaarachchige, Max Hirschberger, N. P. Ong, and R. J. Cava, *Nature* **514**, 205 (2014).

[4] J. T. Ye, Y. J. Zhang, R. Akashi, M. S. Bahramy, R. Arita, and Y. Iwasa, *Science* **338**, 1193 (2012).

[5] Defen Kang, Yazhou Zhou, Wei Yi, Chongli Yang, Jing Guo, Youguo Shi, Shan Zhang, Zhe Wang, Chao Zhang, Sheng Jiang, Aiguo Li, Ke Yang, Qi Wu, Guangming Zhang, Liling Sun, and Zhongxian Zhao, *Nature Commun.* **6**, 6088 (2015).

[6] Xing-Chen Pan, Xuliang Chen, Huimei Liu, Yanqing Feng, Zhongxia Wei, Yonghui Zhou, Zhenhua Chi, Li Pi, Fei Yen, Fengqi Song, Xianggang Wan, Zhaorong Yang, Baigeng Wang, Guanhou Wang, and Yuheng Zhang, *Nature Commun.* **6**, 7805 (2015).

[7] Yan Sun, Shu-Chun Wu, Mazhar N. Ali, Claudia Felser, and Binghai Yan, Preprient at http://arXiv:1508.03501v2 (2015).

[8] Yanpeng Qi, Pavel G. Naumov, Mazhar N. Ali, Catherine R. Rajamathi, Oleg Barkalov, Yan Sun, Chandra Shekhar, Shu-Chun Wu, Vicky Süβ, Marcus Schmidt, EchhardPippel, Peter Werner, ReinaldHillebrand, Tobias Förster, Erik Kampertt, Walter Schnelle, Stuart Parkin, R. J. Cava, Claudia Felser, Binghai Yan, Sergiy A. Medvedev, Preprint at http://arXiv:1508.03502 (2015).

[9] X. Qian, J. Liu, L. Fu, and J. Li, *Science* **346**, 1344 (2014).

[10] D. E. Moncton, J. D. Axe, and F. J. DiSalvo, *Phys. Rev. B* **16**, 801 (1977).

[11] Alexey A Soluyanov, Dominik Gresch, Zhijun Wang, Quansheng Wu, Matthias Troyer, Xi Dai, and B. Andrei Bernevig, http://arXiv: 1507.01603v1(2015).

[12] H. P. Hughes and R. H. Friend, *Journal of Physics C: Solid State Physics* **11**, L103 (1978);

Thorsten Zandt, HelmutDwelk, ChristophJanowitz, and RecardoManzke, *Journal of Alloys and Compounds* **442**, 216 (2007).

[13] Yan Sun, Shu-Chun Wu, Mazhar N. Ali, Claudia Felser, and Binghai Yan, *Phys. Rev. B* **92**, 161107(2015).

[14] P. Giannozzi, et al., *J. Phys.: Condens. Matter* **21**, 395502 (2009).

[15] K.F. Garrity, J.W. Bennett, K.M. Rabe, and D. Vanderbilt, *Comput. Mater. Sci.* 81, 446 (2014).

[16] S. Grimme, *J. Comp. Chem.* **27**, 1787 (2006).

[17] V. Barone et al., *J. Comp. Chem.* **30**, 934 (2009).

[18] H. J. Monkhorst and J. D. Pack, *Phys. Rev. B* 13, 5188 (1976).

[19] H. Padamsee, J. E. Neighbor, and C. A. Shiffman, *J. Low Temp. Phys.* 12, 387 (1973).

[20] W. L. McMillan, *Phys. Rev.* 167, 331 (1968).

[21]Michaela Riflikova, Roman Martoňák, and ErioTosatti, *Phys. Rev. B* 90, 035108 (2014).

[22]L. N. Cooper, *Phys. Rev.* 104, 1189 (1956).

[23] J. Bardeen, L. N. Cooper, and J. R. Schrieffer, *Phys. Rev.* 106, 162(1957).

[24] J. Bardeen, L. N. Cooper, and J. R. Schrieffer, *Phys. Rev.* 108, 1175 (1957).

**Table I:** The physical properties of MoTe$_2$ and MoTe$_{1.8}$S$_{0.2}$ single crystals.

| Parameters | Units | MoTe$_2$ | MoTe$_{1.8}$S$_{0.2}$ |
|---|---|---|---|
| $T_C$ | K | 0.1 | 1.3 |
| $\gamma$ | mJ mol$^{-1}$K$^{-2}$ | 3.06 | 2.07 |
| $\beta$ | mJ mol$^{-1}$K$^{-4}$ | 0.758 | 0.635 |
| $N(E_F)$ (exp.) | states/eV | 1.3 | 0.88 |
| $\Theta_D$ | K | 135 | 143 |
| $\lambda_{ep}$ | | 0.32 | 0.49 |

**Figure 1:**

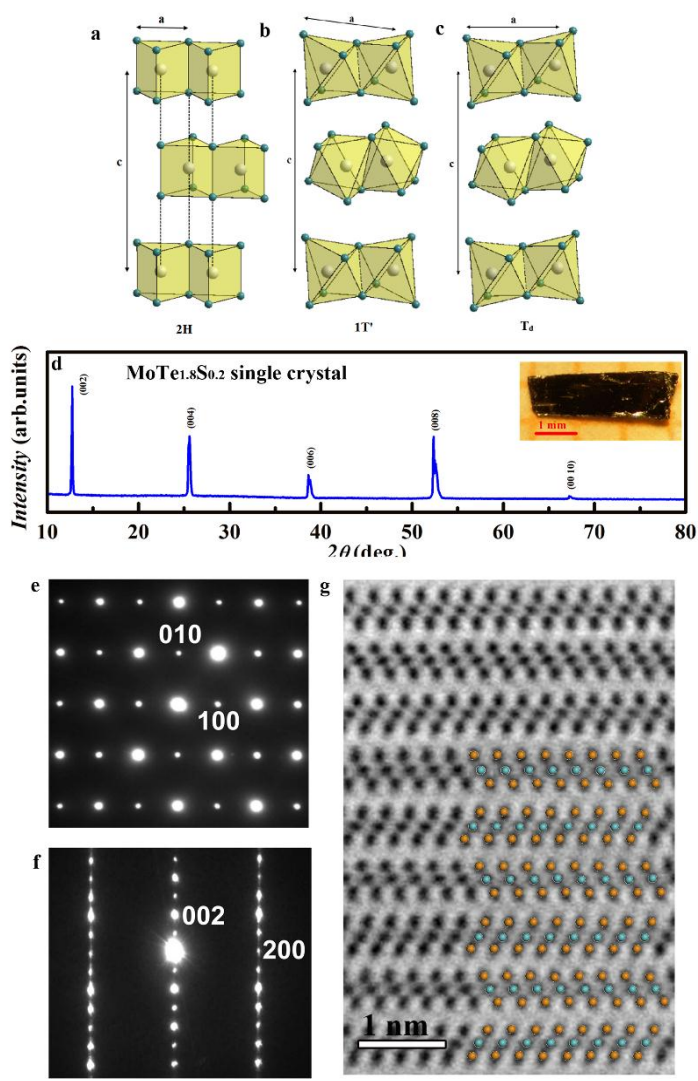

**Fig. 1:** **(a), (b)** and **(c)**: The crystalline structure of *2H*, *1T'* and *$T_d$* structures of MoTe$_2$; **(d)**: XRD patterns of the MoTe$_{1.8}$S$_{0.2}$ single crystal measured on the (00l) surface. Inset presents the picture of the single crystal used for this study. The size is approximately 3*1*0.5; **(e)** and **(f)**: The electron diffraction patterns of MoTe$_{1.8}$S$_{0.2}$ single crystal taken along [001] and [010] zone axis directions, respectively; **(g)** The ABF-STEM image of MoTe$_{1.8}$S$_{0.2}$ single crystal taken along [010] zone direction, the orange and cyan spheres represent the Te/S ions and Mo ones, respectively. Scale bar is 1nm.

**Figure 2:**

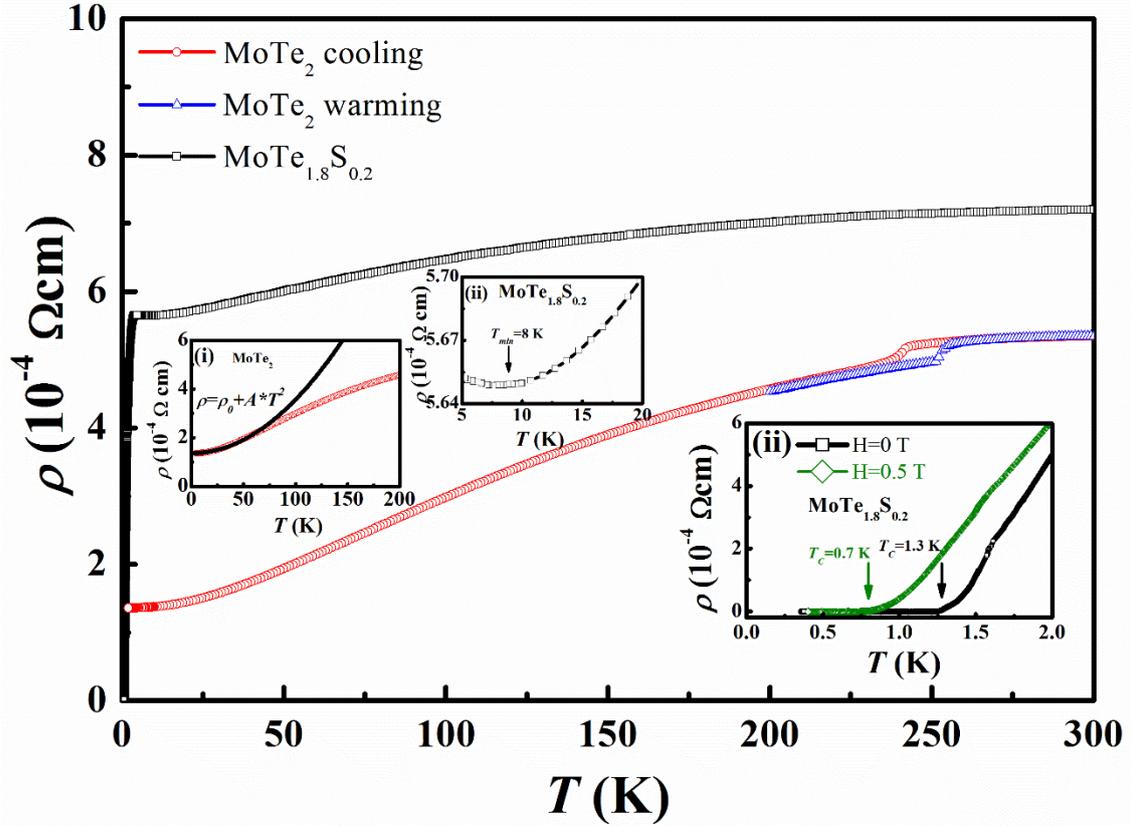

**Fig. 2:** The temperature dependence of resistivity of MoTe$_2$ (warming and cooling modes) and MoTe$_{1.8}$S$_{0.2}$ single crystals. The left inset presents the fitting result according to the Fermi liquid theory. The middle inset shows the minimum of the resistivity $\rho_{min}$ with a larger scale. The right inset presents the temperature dependence of resistivity of MoTe$_{1.8}$S$_{0.2}$ single crystal under H=0 T and 0.5 T at the low temperature.

**Figure 3:**

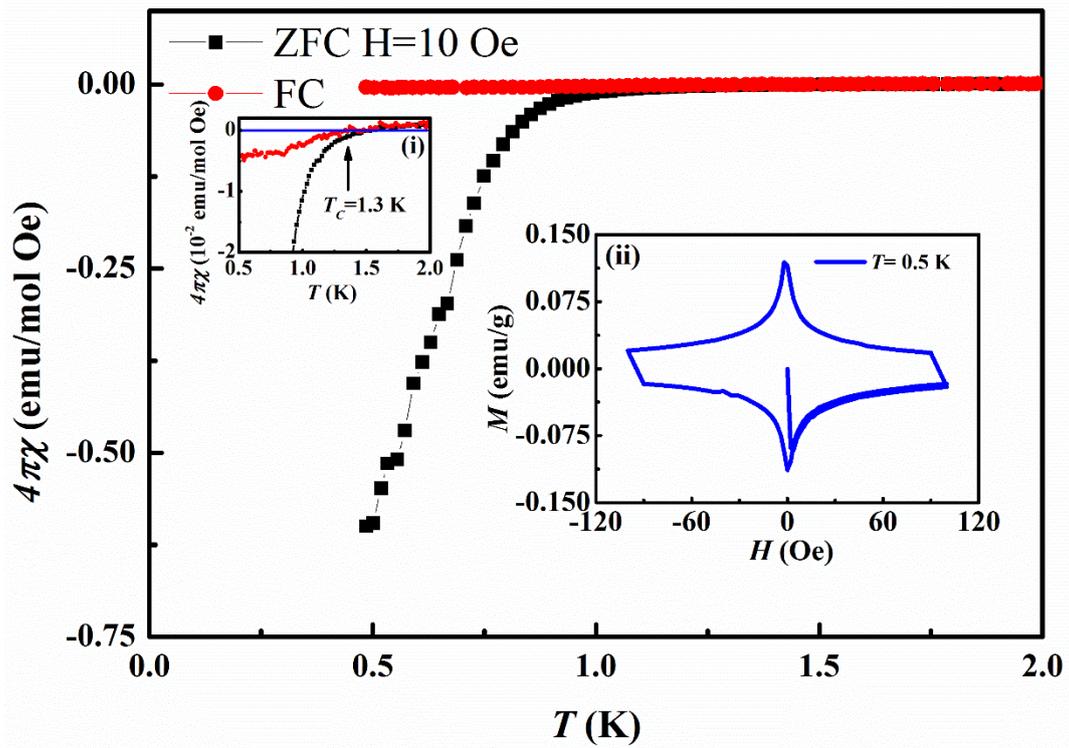

**Fig. 3:** The temperature dependence of magnetization of MoTe$_{1.8}$S$_{0.2}$ single crystal with ZFC and FC modes under H=10 Oe. The left inset presents the temperature dependence of magnetization under a large scale around the $T_C$. The right inset shows the magnetic field dependence of magnetization at $T$=0.5 K.

**Figure 4:**

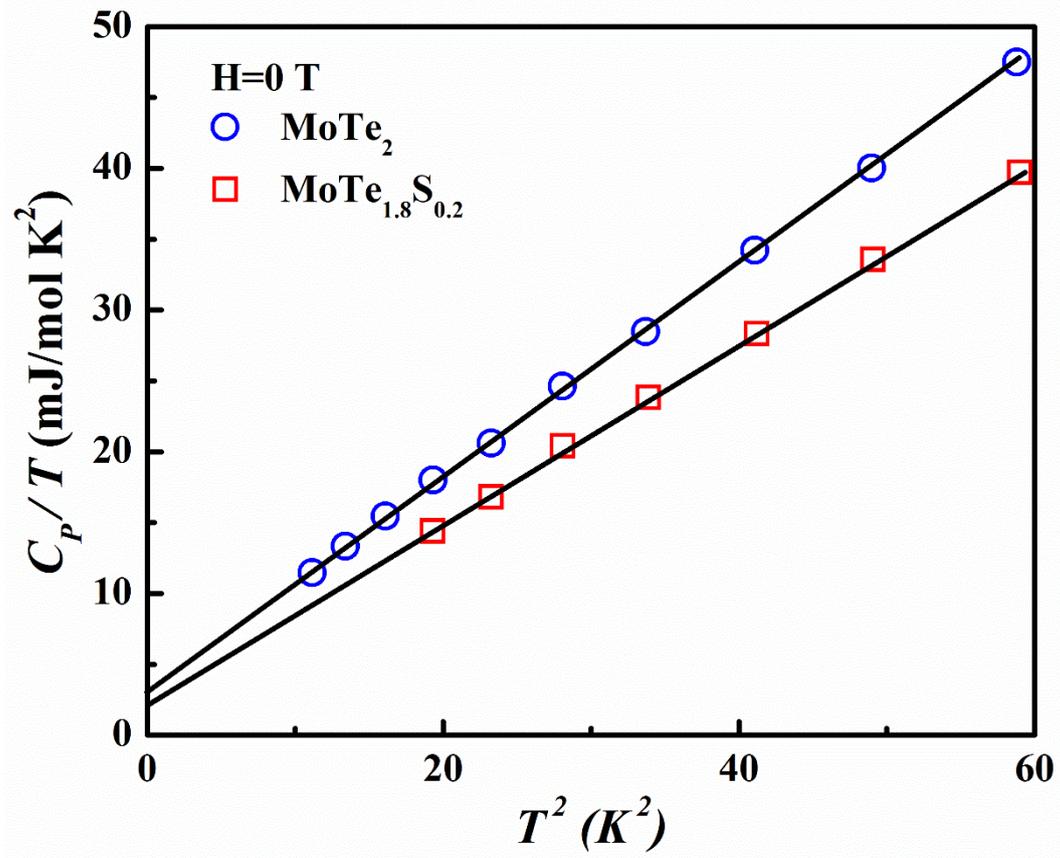

**Fig. 4 :** $T^2$ dependence of $C_P/T$ of MoTe$_2$ and MoTe$_{1.8}$S$_{0.2}$ crystals under zero field. The solid lines show the heat capacity data fitting with the equation $C_P/T=\gamma_N+\beta T^2$.

**Figure 5:**

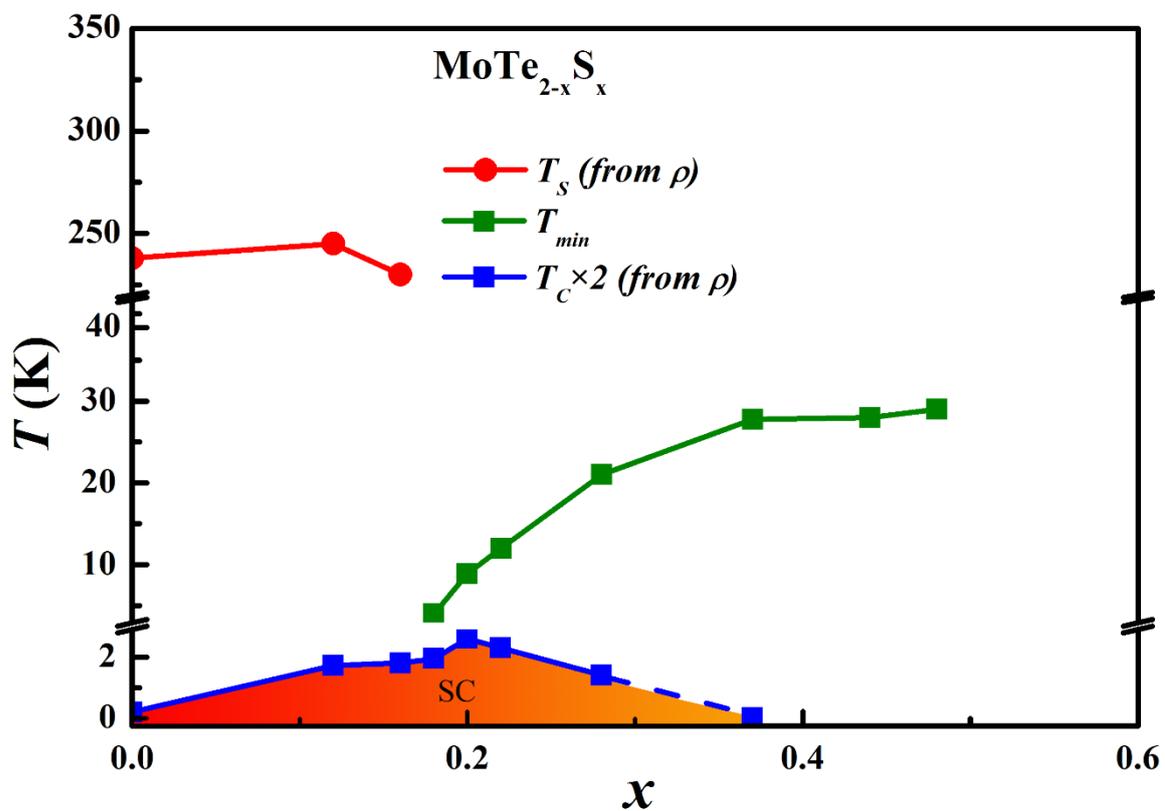

**Fig. 5:** The superconducting phase diagram of MoTe$_{2-x}$S$_x$ single crystals. SC presents the superconductivity.